\documentclass[aps,prb,pdflatex,reprint,showpacs,superscriptaddress]{revtex4-2}
\usepackage{amsmath}
\usepackage{amssymb}
\usepackage{amsthm}
\usepackage{bbm}
\usepackage{graphicx}
\usepackage{hyperref}
\usepackage{physics}
\usepackage{systeme}
\usepackage{times}
\usepackage{epstopdf}
\usepackage{float}

\hypersetup{
  colorlinks=true,linkcolor=blue,citecolor=blue,
  filecolor=blue,urlcolor=blue,breaklinks=true
}

\begin{document}

\title{Relativistic quantum motion of an electron in spinning cosmic string spacetime in the presence of uniform magnetic field and Aharonov-Bohm potential}

\author{M\'{a}rcio M. Cunha}
\email{marciomc05@gmail.com}
\affiliation{
        Departamento de F\'{i}sica,
        Universidade Federal do Maranh\~{a}o,
        65085-580, S\~{a}o Lu\'{i}s, Maranh\~{a}o, Brazil
      }
\author{Edilberto O. Silva}
\email{edilbertoo@gmail.com}
\affiliation{
        Departamento de F\'{i}sica,
        Universidade Federal do Maranh\~{a}o,
        65085-580, S\~{a}o Lu\'{i}s, Maranh\~{a}o, Brazil
      }
\date{\today }
\begin{abstract}
In this manuscript, we study the relativistic quantum mechanics of an electron in external fields in the spinning cosmic string spacetime. We obtain the Dirac equation, write the first and second-order equations from it, and then we solve these equations for bound states. We show that there are bound states solutions for the first-order equation Dirac. For the second-order equation, we show that its wave functions are given in terms of the Kummer functions and we determine the energies of the particle. We examine the behavior of the energies as a function of the physical parameters of the model, such as rotation, curvature, magnetic field, Aharonov-Bohm flux, and quantum numbers. Our study reveals that both curvature and rotation influence more intensely when these parameters have values smaller than 0.3. We also find that, depending on the values of these parameters, there are energy nonpermissible levels.
\end{abstract}

\maketitle

\section{Introduction}
\label{intro}

Symmetry is key-ingredient in the description of natural phenomena. The notion of symmetry is an essential feature in several areas of physics.  In this context, the well-known Noether's theorem \cite{lanczos2012variational} establishes a connection between symmetry and conservation laws of relevant physical quantities. In quantum mechanics, we often use symmetry to obtain crucial results concerning angular momenta operators \cite{sakurai2014modern}.
Likewise, symmetry is important in the topic of quantum information \cite{PhysRevB.95.045111}.
Symmetry is also important in the framework of relativity \cite{westman2009coordinates}, and for this reason, it is indispensable in research areas such as particle physics \cite{gibson1980symmetry} and cosmology \cite{preskill1991cosmology}.

A pertinent question in the research areas cited above refers to think about what happens when some symmetry is broken in a given physical system.
When a system suffers a phase transition, for example, it can lose some type of symmetry.
Another example of symmetry-breaking occurs in condensed matter systems: by employing the Volterra process \cite{solyom2007fundamentals}, it is possible to break some symmetry of the system due to the creation of a topological defect, like disclinations and dislocations \cite{RevModPhys.80.61,puntigam1997volterra}, for instance.

Topological defects can emerge in a large number of physical systems covering themes such as liquid crystals \cite{lavrentovich2012defects}, graphene physics \cite{alden2013strain}, magnetism \cite{kou2011tunable} and cosmology \cite{durrer1999topological}.
Recent studies also have reported the importance of topological defects in Life Sciences \cite{kawaguchi2017topological,saw2017topological}.
On the point of view of cosmology, defects in the spacetime topology can be viewed as a possible consequence of the evolution of the early universe, which has suffered phase transitions due to the temperature decreasing and the process of expansion \cite{hindmarsh1995cosmic,vilenkin2000cosmic}.

 In this contribution, we are particularly involved in studying the topological defect known as a cosmic string. A cosmic string is a linear defect, similar to a flux tube in type-II superconductors \cite{hindmarsh1995cosmic}. The spacetime around a cosmic string has a conical symmetry, identically to the case of a disclination \cite{moraes2000condensed}. The concept of a cosmic string it was introduced in the literature by Kibble \cite{kibble1976topology}. Since then, this topic has been investigated in diverse forms.  An intriguing facet in this subject refers to the quantum mechanical description of a particle in a region of the spacetime containing a cosmic string. It can be done both in the scenario of relativistic and nonrelativistic quantum mechanics. For instance, the hydrogen atom in a spacetime of a cosmic string it was analyzed in Ref. \cite{PhysRevD.66.105011}. In Ref. \cite{marques2005exact}, it was considered the problem of a relativistic electron in the presence of both Coulomb and scalar potentials in the cosmic string spacetime. Results about vacuum polarization in a cosmic string spacetime were reported in Ref. \cite{PhysRevD.74.025017}.
Again, the cosmic string spacetime it was considered as a background to examine relativistic oscillators \cite{PhysRevA.84.032109}, quantum phases \cite{bakke2016relativistic}, and fermionic currents \cite{bezerra2016induced}.

 A relevant issue in this context consists of taking into consideration the influence of electromagnetic fields in the quantum particle motion. Landau levels \cite{Book.2005.Griffiths} and the Aharonov-Bohm effect \cite{PR.1959.115.485,peshkin1989aharonov}, for instance, are essential ingredients in the investigation of quantum systems even in a flat spacetime. It can be explained because Landau levels are a quantum analog of classical cyclotron motion, while the Aharonov-Bohm effect reveals the significance of the vector potential in the quantum world.
Then, studying the contribution of magnetic fields to the quantum mechanical description of a system in spacetime having a topological defect is a natural development. Examples of studies dealing with Landau levels and the Aharonov-Bohm effect in the presence of topological defects can be accessed in Refs. \cite{de2001landau} and \cite{azevedo1998topological}, respectively. In particular, the inclusion of electromagnetic interactions in the case of a cosmic string background also has been considered. For instance, in Ref. \cite{medeiros2012relativistic}, it was analyzed the quantum dynamics of a charged particle in the presence of a magnetic field and scalar potential. In Ref. \cite{PhysRevD.93.043545}, various configurations of confined magnetic fields are examined and the existence of induced vacuum fermionic currents is investigated.

On the other hand, we can be interested in analyzing the behavior of a quantum system when noninertial effects turn on. These effects play a fundamental role in the description of systems governed by classical mechanics. Noninertial effects also can take place on quantum systems, providing novel theoretical predictions and feasible experimental developments.
For instance, the emergence of quantum phases in rotating systems, in analogy to the Aharonov-Bohm effect \cite{aharonov1973quantum,semon1982experimental} were investigated. In addition, a relation between the Hall effect and inertial forces it was established \cite{johnson2000inertial}.
Besides, if a given system it is put to rotate, it has consequences in diverse physical properties like spin transport \cite{PhysRevB.84.104410,chowdhury2014spin}, electronic structure \cite{garcia2017geometric}, and even can present magnetization due rotation, like in the Barnett effect \cite{PhysRevB.92.174424}. While a magnetic field produces a spin-field coupling, resulting in the anomalous
Zeeman effect \cite{nouredine2009quantum}, rotation produces an analog effect, due the spin-rotation coupling \cite{danner2020spin}.
Thus, rotation can contribute similarly to a magnetic field in the dynamics of a quantum system.
More, noninertial effects are an interesting issue in the situation in which spacetime contains topological defects. In this case, the noninertial effects and the presence of a topological defect can be included in the quantum mechanical description by employing the same tools: we can use a metric tensor to a spinning spacetime with a topological defect \cite{clement1990rotating}.
The spacetime of a spinning cosmic string has been considered as background for several problems involving quantum systems. For instance, the Schr\"{o}dinger equation in that spacetime it was solved in Ref. \cite{hassanabadi2015motion}. Bound states for neutral particles in a rotating frame of a cosmic string were analyzed in Ref. \cite{PhysRevD.82.084025}. Likewise, rotating effects on a Landau-Aharonov-Casher System in the spacetime of a cosmic string were investigated in Ref. \cite{bakke2015rotating}.

As we already have mentioned, in some cases rotation presents similarities within electromagnetic fields. This way, it is also an attractive question examining how the electromagnetic interactions affect the particle quantum motion of a rotating system in the presence of a topological defect. A recent example of studying dealing with both topological and noninertial effects in the presence of an Aharonov-Bohm potential can be accessed in Ref. \cite{oliveira2019topological}. In Ref. \cite{Wang_canadian_journal2017}, it was addressed the problem of a spinless relativistic particle subjected to a uniform magnetic field in the spinning cosmic string spacetime.
The Dirac oscillator in the spacetime of a cosmic string considering noninertial effects and the presence of the  Aharonov–Casher effect it was analyzed in Ref. \cite{EPJC_oliveira2019topological}.
In Ref. \cite{wang2018study}, it was analyzed the problem of a charged half-spin particle depicted by the Dirac equation in the presence of a uniform magnetic field in the rotating cosmic string spacetime.
A meaningful aspect in this context consists of analyzing how different configurations of magnetic fields affect the quantum particle motion.

In this paper, we study the relativistic quantum mechanics of an electron in the presence of both a uniform magnetic field and Aharonov-Bohm potential in the spinning cosmic string spacetime. In other words, we solve the Dirac equation in this scenario and investigate how the rotation, curvature and external magnetic fields affect the wave functions and energies of the electron.

The manuscript is organized as follows. In Section II, we present some algebraic elements necessary to construct the field equations in curved spacetime and write the Dirac equation describing the quantum motion of the electron in the presence of external magnetic fields in the spinning cosmic string background. In Section III, we deal with first-order solutions and study the existence of isolated solutions for the particular case of a particle at rest. In Section IV, we take our attention to the case when the energy of the particle is different from its rest energy. We map the Dirac equation problem in curved space with minimal coupling into a Sturm-Liouville problem for the upper component of the Dirac spinor and, using an appropriate ansatz, we derive the radial equation. We solve the radial equation and find the wave functions and energies of the particle. We make a detailed discussion of the results and also comparisons with other studies in the literature. In Section V, we present our conclusions. In our work, we use natural units, $\hbar = c = G = 1$.

\section{Dirac equation in the spinning cosmic string spacetime}
In this section, we briefly present the main tools needed to construct the Dirac equation in the conical spacetime in the presence of noninertial effects. The first step consists in take a look at the metric tensor characterizing this geometry. Next, we will choose an appropriate tetrad basis and implement the fields configuration involved through the performing of a minimal substitution.
The spacetime induced by a rotating cosmic string is described by the metric
\begin{equation}
ds^{2}=\left( dt+ad\varphi \right) ^{2}-dr^{2}-\alpha ^{2}r^{2}d\varphi
^{2}-dz^{2},  \label{metric}
\end{equation}
where $-\infty <z<\infty $, $r\geqslant 0$ and $0\leqslant \varphi \leqslant
2\pi$. The parameter $\alpha $ is related to the linear mass density $\mu$
of the cosmic string through the relation $\alpha =1-4\mu$ and it runs in
the interval $(0,1]$. The quantity $a=4J$ is the rotation parameter, with $J$ representing the angular momentum of the spinning cosmic string. The relativistic quantum dynamics of a spin-$1/2$ particle interacting with external magnetic fields in the rotating cosmic string spacetime is governed by the Dirac equation
\begin{equation}
\left[ i\gamma ^{\mu }\left( x\right) \left( \partial _{\mu }+\Gamma _{\mu
}\left( x\right) +ieA_{\mu }(x)\right) -M\right] \Psi \left( x\right) =0,
\label{diracsc}
\end{equation}
where $M$ is the mass of the particle and $\gamma ^{\mu }\left( x\right)$ are the Dirac matrices in the
rotating cosmic string spacetime, which are defined in terms of the tetrad fields $e_{a}^{\mu }$ and Dirac matrices in the flat space $\gamma ^{a}$ in the following way:
\begin{equation}
\gamma ^{\mu }\left( x\right) =e_{a}^{\mu }\left( x\right) \gamma ^{a},
\label{gmatrices}
\end{equation}
where
\begin{equation}
\gamma ^{a}=\left( \gamma ^{0},\gamma ^{i}\right),\, \text{with}\, \gamma ^{0}=\left(
\begin{array}{cc}
1 & 0 \\
0 & -1
\end{array}
\right),\,\gamma ^{i}=\left(
\begin{array}{cc}
0 & \sigma ^{i} \\
-\sigma ^{i} & 0
\end{array}
\right),
\end{equation}
are the standard Dirac matrices
and $\sigma ^{i}=\left( \sigma ^{x},\sigma ^{y},\sigma ^{z}\right)$ are the usual Pauli matrices. The matrices (\ref{gmatrices}) satisfy the following relation:
\begin{equation}
\left\{ \gamma ^{\mu }\left( x\right) ,\gamma ^{\nu }\left( x\right)
\right\} =2g^{\mu \nu }\left( x\right).
\end{equation}
Also, in Eq. (\ref{diracsc}), $\Gamma _{\mu }\left( x\right)$ is the
spin affine connection given by
\begin{equation}
\Gamma _{\mu }\left( x\right) =\frac{1}{4}\gamma ^{a}\gamma ^{b}e_{a}^{\nu
}\left( x\right) \left[ \partial _{\mu }e_{b\nu }\left( x\right) -\Gamma
_{\mu \nu }^{\sigma }e_{b\sigma }\left( x\right) \right], \label{affine}
\end{equation}
where $\Gamma_{\mu \nu}^{\sigma}$ are the Christoffel symbols of the second kind and $e_{a}^{\mu }(x)$ is the tetrad field. The tetrad basis satisfies the relations
\begin{align}
&e_{\mu }^{a}\left( x\right) e_{\nu}^{b}\left(x\right) \eta_{ab}=g_{\mu \nu }\left( x\right),\\
&e_{\mu}^{a}\left(x\right) e_{\nu
}^{b}\left(x\right)=\delta_{a}^{b},\\
&e_{a}^{\mu}\left(x\right)
e_{\nu}^{a}\left(x\right) =\delta_{\mu}^{\nu}.
\end{align}
In Eq. (\ref{affine}), the Greek letters are used for tensor indices while the Latin letters are denoting Minkowski indices. We use the tetrad basis and its inverse defined as \cite{EPJC.2019.79.311}
\begin{eqnarray}
e_{\mu }^{a}\left( x\right)  &=&\left(
\begin{array}{cccc}
1 & 0 & a & 0 \\
0 & \cos \varphi  & -r\alpha \sin \varphi  & 0 \\
0 & \sin \varphi  & r\alpha \cos \varphi  & 0 \\
0 & 0 & 0 & 1
\end{array}
\right) , \\
e_{a}^{\mu }\left( x\right)  &=&\left(
\begin{array}{cccc}
1 & \frac{a\sin \varphi }{r\alpha } & -\frac{a\cos \varphi }{r\alpha } & 0
\\
0 & \cos \varphi  & \sin \varphi  & 0 \\
0 & -\frac{\sin \varphi }{r\alpha } & \frac{\cos \varphi }{r\alpha } & 0 \\
0 & 0 & 0 & 1
\end{array}
\right).\label{trdi}
\end{eqnarray}
For this choice, it can be shown that the non-vanishing affine connection is given by
\begin{equation}
\Gamma _{\mu }=\left( 0,0,\Gamma _{\varphi },0\right), \, \text{with} \; \Gamma _{\varphi }=\frac{i}{2}\left( 1-\alpha \right) \Sigma^{z},
\end{equation}
with
\begin{equation}
\Gamma _{\varphi }=\frac{i}{2}\left( 1-\alpha \right) \Sigma^{z},
\end{equation}
where
\begin{equation}
\Sigma ^{z}=\left(
\begin{array}{cc}
\sigma ^{z} & 0 \\
0 & \sigma ^{z}
\end{array}
\right), \;\; \sigma ^{z} =\left(
\begin{array}{cc}
1 & 0 \\
0 & -1%
\end{array}
\right).
\end{equation}
By using the tetrad basis (\ref{trdi}), the matrices (\ref{gmatrices}) can be written explicitly as
\begin{align}
&\gamma ^{t} =\gamma ^{0}-a\gamma ^{\varphi },  \\
&\gamma ^{r}=\left(
\begin{array}{cc}
0 & \sigma ^{r} \\
-\sigma ^{r} & 0
\end{array}
\right), \;\;\gamma ^{\varphi }=\left(
\begin{array}{cc}
0 & \sigma ^{\varphi } \\
-\sigma ^{\varphi } & 0
\end{array}
\right),
\end{align}
with
\begin{equation}
\sigma^{r} =\left(
\begin{array}{cc}
0 & e^{-i\varphi } \\
e^{+i\varphi } & 0
\end{array}
\right), \;\;\sigma ^{\varphi }=\frac{1}{r\alpha }\left(
\begin{array}{cc}
0 & -ie^{-i\varphi } \\
ie^{+i\varphi } & 0
\end{array}
\right)
\end{equation}
being the Pauli matrices in the curved spacetime.

Since we are first interested in studying the solutions of the Dirac equation in its present form (Eq. (\ref{diracsc})), we need to write the corresponding system of first order coupled differential equations. For this to be accomplished, let's assume the time-dependence of the wave functions together with the decomposition of the fermion field in the form
\begin{equation}
    \Psi \left( r,\varphi \right)=e^{-iEt}\left(
\begin{array}{c}
\psi _{1}\left( r,\varphi \right)  \\
\psi _{2}\left( r,\varphi \right)
\end{array}
\right),\label{S1}
\end{equation}
with
\begin{align}
&\psi _{1}\left( r,\varphi \right)=\left(
\begin{array}{c}
\psi _{a}\left( r,\varphi \right)  \\
\psi _{b}\left( r,\varphi \right)
\end{array}
\right) =\left(
\begin{array}{c}
e^{im\varphi }f_{+}\left( r\right)  \\
ie^{i\left( m+1\right) \varphi }f_{-}\left( r\right)
\end{array}
\right) ,\label{P1} \\
& \psi _{2}\left( r,\varphi \right) =\left(
\begin{array}{c}
\psi _{c}\left( r,\varphi \right)  \\
\psi _{d}\left( r,\varphi \right)
\end{array}
\right) =\left(
\begin{array}{c}
e^{im\varphi }g_{+}\left( r\right)  \\
ie^{i\left( m+1\right) \varphi }g_{-}\left( r\right)
\end{array}
\right). \label{P2}
\end{align}
The system we will analyze takes into account the particle is immersed in a region where there is a uniform magnetic field and also the potential due to a thin long solenoid along the z-axis. Having this field configuration in mind, we study
the physical implications due to noninertial effects and the Aharonov-Bohm potential on the relativistic Landau quantization. We also take into account the translational invariance of the system along the $z$-direction, which allows us to eliminate the third direction ($p_z=z=0$) and, consequently, we can consider only the planar motion \cite{PRL.1990.64.503,PRD.1994.50.7715,PRD.2012.85.041701,AoP.2013.339.510}. In this case, the
particle experiences a superposition of potential vectors written in the
Coulomb gauge as
\begin{equation}
\mathbf{A}=\left( 0,-\alpha rA_{\varphi }, 0\right) ,  \label{vectorA}
\end{equation}
with
\begin{align}
A_{\varphi } &=A_{\varphi ,1}+A_{\varphi ,2} , \label{potential} \\
A_{\varphi ,1} &=\frac{Br}{2},\;\;\; A_{\varphi ,2}=\frac{\phi }{\alpha r},\label{potentialv}
\end{align}
where $B$ is the magnetic field magnitude, $\phi =\Phi /\Phi _{0}$, $\Phi$ is the
magnetic flux and $\Phi _{0}=2\pi /e$ is the quantum of magnetic flux along the solenoid. This configuration also provides an superposition of magnetic fields in the
z-direction
\begin{equation}
B=B_{z,1}+B_{z,2},
\end{equation}
with
\begin{equation}
B_{1,z}=B,  \;\;\; B_{z,2}=\phi \frac{\delta (r)}{\alpha r},  \label{fields}
\end{equation}
Note that the particle only interacts with the magnetic field due to the
potential vector $A_{\varphi ,1}$. Here, we are focused on studying the
electron motion only in the $r\neq 0$ region, so that we can neglect the
point interaction $B_{z,2}$ and, consequently, consider only regular wave
functions.

Using the results above, the Dirac equation (\ref{diracsc}) can be written as
\begin{align}
(E-&M) \,\psi_{1}+\sigma ^{r}i\partial _{r}\psi _{2}\notag\\ &+\sigma
^{\varphi }\left( i\partial _{\varphi }+eA_{\varphi }-aE-\frac{s}{2}\left(
1-\alpha \right) \right) \psi_{2}=0, \label{Eqpsi}\\
(E+&M) \,\psi_{2}+\sigma ^{i}i\partial _{r}\psi _{1}\notag \\ &+\sigma
^{\varphi }\left( i\partial _{\varphi }+eA_{\varphi }-aE-\frac{s}{2}\left(
1-\alpha \right) \right) \psi_{1}=0. \label{Eqchi}
\end{align}
At this point, we are ready to solve the equations (\ref{Eqpsi}) and (\ref{Eqchi}) by considering two distinct circumstances:

(i) Take our attention to isolated solutions of the first order Dirac equation by imposing the condition $E= \pm M$;

(ii) By imposing the condition $E\neq \pm M$,  we looking for solutions of the second order Dirac equation.

We will show in the next two sections that there are bound state solutions for both cases and discuss their main physical properties. To distinguish each case in (i), in the next section we use the superscripts ($\pm$) to label the quantities corresponding to $E= \pm M$.

\section{Solution of the equation of motion to $E=\pm M$}

To study the existence of isolated solutions of the Dirac equation (\ref{diracsc}), we must set $E=\pm M$ in Eqs.  (\ref{Eqpsi}) and (\ref{Eqchi}). In literature, such solutions are known to be excluded from the Sturm-Liouville problem. The search for isolated solutions of the Dirac equation has been performed in different physical contexts \cite
{EPJC.2019.79.596,EPL.2014.108.30003,AoP.2013.338.278,JPA.2007.40.263,PLA.2006.351.379}. The bound state solution must satisfy the normalization condition
\begin{equation}
\int_{0}^{\infty }\left( |\psi_{1}(r)|^{2}+|\psi_{2}(r)|^{2}\right) rdr=1.
\label{norm}
\end{equation}
By making $E=+M$ in Eqs. (\ref{Eqpsi}) and (\ref{Eqchi}) and using Eqs. (\ref{P1}) and (\ref{P2}), we get
\begin{align}
&\frac{dg_{+}^{(+)}(r)}{dr}-\frac{L_{m}^{(+)}}{r\alpha }g_{+}^{(+)}(r)+\frac{eBr}{2}g_{+}^{(+)}(r) =0,\label{eqg1}
\\
&\frac{dg_{-}^{(+)}(r)}{dr}+\frac{L_{m+1}^{(+)}}{r\alpha }g_{-}^{(+)}(r)-\frac{eBr}{2}g_{-}^{(+)}(r) =0,\label{eqg2}
\\
&\frac{df_{+}^{(+)}(r)}{dr}-\frac{L_{m}^{(+)}}{r\alpha }f_{+}^{(+)}(r)+\frac{eBr}{2}f_{+}^{(+)}(r)
=-2Mg_{-}^{(+)}(r), \label{eqg3}\\
&\frac{df_{-}^{(+)}(r)}{dr}+\frac{L_{m+1}^{(+)}}{r\alpha }f_{-}^{(+)}(r)-\frac{eBr}{2}f_{-}^{(+)}(r)
=2Mg_{+}^{(+)}(r),\label{eqg4}
\end{align}
with
\begin{align}
L_{m}^{(+)}& =m-\phi +aM+\frac{s}{2}\left( 1-\alpha \right),  \label{lmm1}\\
L_{m+1}^{(+)}& =m+1-\phi +aM+\frac{s}{2}\left( 1-\alpha \right).\label{lmm2}
\end{align}
The solution of the coupled linear differential equations system (\ref{eqg1})-(\ref{eqg4}) is given by
\begin{align}
f_{+}^{(+)}(r)& =e^{-\frac{1}{4}eBr^{2}}r^{\frac{L_{m}^{(+)}}{\alpha }}\left[
a_{2}+a_{1}M\left( -\frac{eB}{2}\right) ^{\Omega _{a}}\Gamma _{a}^{(+)}\right],\label{sp1} \\
f_{-}^{(+)}(r)& =e^{\frac{1}{4}eBr^{2}}r^{-\frac{L_{m+1}^{(+)}}{\alpha }}\left[
b_{2}-b_{1}M\left( \frac{eB}{2}\right) ^{-\Omega _{b}}\Gamma _{b}^{(+)}\right] , \label{sp2} \\
g_{+}^{(+)}(r)& =b_{1}e^{-\frac{1}{4}Ber^{2}}r^{\frac{L_{m}^{(+)}}{\alpha }}, \label{sp3} \\
g_{-}^{(+)}(r)& =a_{1}e^{\frac{1}{4}Ber^{2}}r^{-\frac{L_{m+1}^{(+)}}{\alpha }},\label{sp4}
\end{align}
with
\begin{align}
\Omega _{a}& =\frac{1}{2\alpha }\left( L_{m}^{(+)}+L_{m+1}^{(+)}-\alpha
\right) , \\
\Omega _{b}& =\frac{1}{2\alpha }\left( L_{m}^{(+)}+L_{m+1}^{(+)}+\alpha
\right) ,
\end{align}
where
\begin{align}
\Gamma _{a}^{(+)}& =\Gamma \left( -\Omega _{a},-\frac{1}{2}eBr^{2}\right) , \\
\Gamma _{b}^{(+)}& =\Gamma \left( \Omega _{b},\frac{1}{2}eBr^{2}\right) .
\end{align}
are upper incomplete Gamma functions \cite{Book.1972.Abramowitz}, and
$a_{1}$, $a_{2}$, $b_{1}$ and $b_{2}$ are constants. Analyzing the solutions (\ref{sp1}) and (\ref{sp3}), we note that
$e^{-\frac{1}{4}eBr^{2}}$ dominates over $r^{\frac{L_{m}^{(+)}}{\alpha }}$
for any value of $L_{m}^{(+)}/\alpha $, in such way both solutions converge when $r\rightarrow 0$ and $r\rightarrow \infty $. This will not occur for the
function $e^{\frac{1}{4}eBr^{2}}$ in the solutions (\ref{sp2}) and (\ref{sp4}). Moreover, since the incomplete Gamma
functions $\Gamma _{a}^{(+)}$ and $\Gamma _{b}^{(+)}$ always diverge, then
the function $f_{+}^{(+)}(r)$ will only converges as $r\rightarrow 0$ if
$a_{1}=0$ while the function $f_{-}^{(+)}(r)$ will always diverge when $r\rightarrow \infty$ and,
therefore, will not be a square-integratable function. Thus, the only solution allowed for the equations system (\ref{eqg1})-(\ref{eqg4}) results
\begin{equation}
f_{+}^{(+)}(r)=a_{2}e^{-\frac{1}{4}eBr^{2}}r^{\frac{L_{m}^{(+)}}{\alpha }},\, \text{with}\;\frac{L_{m}^{(+)}}{\alpha }\geqslant 0 ,\label{fnl}
\end{equation}
with $f_{-}^{(+)}(r)=g_{+}^{(+)}(r)=g_{-}^{(+)}(r)=0$. Solution (\ref{fnl})
satisfies equation (\ref{norm}) and constitutes a bound state solution for the case $
E=M$, i.e., an isolated solution to the Dirac equation (\ref{diracsc}) in the metric spacetime (\ref{metric}).

Proceeding in an analogous way, now we make $E=-M$ in Eqs. (\ref{Eqpsi}) and (\ref{Eqchi}). We find the system of equations
\begin{align}
&\frac{df_{+}^{(-)}(r)}{dr}-\frac{L_{m}^{(-)}}{r\alpha }f_{+}^{(-)}(r)+\frac{eBr}{2}f_{+}^{(-)}(r)=0,  \label{f1}
\\
&\frac{df_{-}^{(-)}(r)}{dr}+\frac{L_{m+1}^{(-)}}{r\alpha }f_{-}^{(-)}(r)-\frac{eBr}{2}f_{-}^{(-)}(r)=0,\label{f2}\\
&\frac{dg_{+}^{(-)}(r)}{dr}-\frac{L_{m}^{(-)}}{r\alpha }g_{+}^{(-)}(r)+\frac{eBr}{2}g_{+}^{(-)}(r)
=2Mf_{-}^{(-)}(r),  \label{g1}\\
&\frac{dg_{-}^{(-)}(r)}{dr}+\frac{L_{m+1}^{(-)}}{r\alpha }g_{-}^{(-)}(r)-\frac{eBr}{2}g_{-}^{(-)}(r)
=-2Mf_{+}^{(-)}(r). \label{g2}
\end{align}
with
\begin{eqnarray}
L_{m}^{(-)} &=&m-\phi -aM+\frac{s}{2}\left( 1-\alpha \right), \label{lpm1} \\
L_{m+1}^{(-)} &=&m+1-\phi -aM+\frac{s}{2}\left( 1-\alpha \right).\label{lpm2}
\end{eqnarray}
The solution of the coupled linear ordinary differential equations system (\ref{f1})-(\ref{g2}) is given by
\begin{align}
f_{+}^{(-)}(r)& =c_{1}e^{-\frac{1}{4}eBr^{2}}r^{\frac{L_{m}^{(-)}}{\alpha }},\label{sms1}
\\
f_{-}^{(-)}(r)& =d_{1}e^{\frac{1}{4}Ber^{2}}r^{-\frac{L_{m+1}^{(-)}}{\alpha }%
}, \label{sms2}\\
g_{+}^{(-)}(r)& =e^{-\frac{1}{4}Ber^{2}}r^{\frac{L_{m}^{(-)}}{\alpha}
}\left[-d_{1}M\left( -\frac{eB}{2}\right) ^{\Lambda _{c}}\Gamma
_{c}^{\left( -\right) }+d_{2}\right], \label{sms3}\\
g_{-}^{(-)}(r)& =e^{\frac{1}{4}eBr^{2}}r^{-\frac{L_{m+1}^{(-)}}{\alpha }}%
\left[ c_{1}M\left( \frac{eB}{2}\right) ^{-\Lambda _{d}}\Gamma _{d}^{\left(
-\right) }+c_{2}\right] \label{sms4},
\end{align}
with
\begin{align}
\Lambda _{c}& =\frac{1}{2\alpha }\left( L_{m}^{(-)}+L_{m+1}^{(-)}-\alpha
\right) , \\
\Lambda _{d}& =\frac{1}{2\alpha }\left( L_{m}^{(-)}+L_{m+1}^{(-)}+\alpha
\right) ,
\end{align}
where
\begin{align}
\Gamma _{c}^{\left( -\right) }& =\Gamma \left( -\Lambda _{c},-\frac{1}{2}
Ber^{2}\right) , \\
\Gamma _{d}^{\left( -\right) }& =\Gamma \left( \Lambda _{d},\frac{1}{2}
eBr^{2}\right) .
\end{align}
By making the same analysis of the solutions as we have made for the case $E=M$, i.e., analyzing the behavior of the functions for $r\rightarrow \pm\, \infty$, we find that the only solution that admits bound state is (\ref{sms3}). Thus, the solution for the case $E=-M$ satisfying the normalization condition (\ref{norm}) is given by
\begin{equation}
g_{+}^{(-)}(r) =d_{2}e^{-\frac{1}{4}Ber^{2}}r^{\frac{L_{m}^{(-)}}{\alpha }},\, \text{with}\;\frac{L_{m}^{(-)}}{\alpha }\geqslant 0,\label{slem}
\end{equation}
with $f_{+}^{(-)}(r) =
f_{-}^{(-)}(r)=g_{-}^{(-)}(r)=0$.
Note that the solutions (\ref{fnl}) and (\ref{slem}) are affected by rotation through Eqs. (\ref{lmm1}) and (\ref{lpm1}), respectively.

\section{Solution of the equation of motion to $E\neq \pm M$}

In this section, we solve the second order equation to $\psi$ that we find from the Eqs. (\ref{Eqpsi}) and (\ref{Eqchi}). The solution of this equation is different from that one calculated in the previous section and allow us to obtain an expression for the particle energies. By isolating $\psi_{2}$ in Eq. (\ref{Eqchi}) and replacing in Eq. (\ref{Eqpsi}), we are able to write the second order differential equation for $\psi_{1}$ as
\begin{align}
&\left(E^{2}-M^{2}\right) \psi_{1} +\partial _{r}^{2}\psi_{1} +\frac{1}{r}\partial
_{r}\psi_{1} +\frac{1}{\alpha r}\sigma ^{z}e\left( \partial _{r}A_{\varphi
}\right) \psi_{1}  \notag \\
& +\frac{1}{\alpha ^{2}r^{2}}\left( \partial _{\varphi }-ieA_{\varphi }+i%
\frac{1-\alpha }{2}\sigma ^{z}+iaE\right) ^{2}\psi_{1} =0.\label{eav}
\end{align}
Using the decomposition of the fermion field (\ref{P1}) (ignoring the subscript ($+$)) together with Eqs. (\ref{vectorA}), (\ref{potential}) and (\ref{potentialv}), we obtain the radial equation for $f(r)$
\begin{equation}
\left( \frac{d^{2}}{dr^{2}}+\frac{1}{r}\frac{d}{dr}-\frac{L^{2}}{\alpha
^{2}r^{2}}-\frac{e^{2}B^{2}r^{2}}{4}+k^{2}\right) f\left( r\right) =0, \label{re}
\end{equation}
where
\begin{equation}
k^{2}=E^{2}-M^{2}+\frac{eB}{\alpha}L+seB,
\end{equation}
\begin{equation}
L=m-\phi +\frac{s\left( 1-\alpha \right) }{2}+aE. \label{am}
\end{equation}
\begin{figure}[!t]
	\centering	
	\includegraphics[width=\columnwidth]{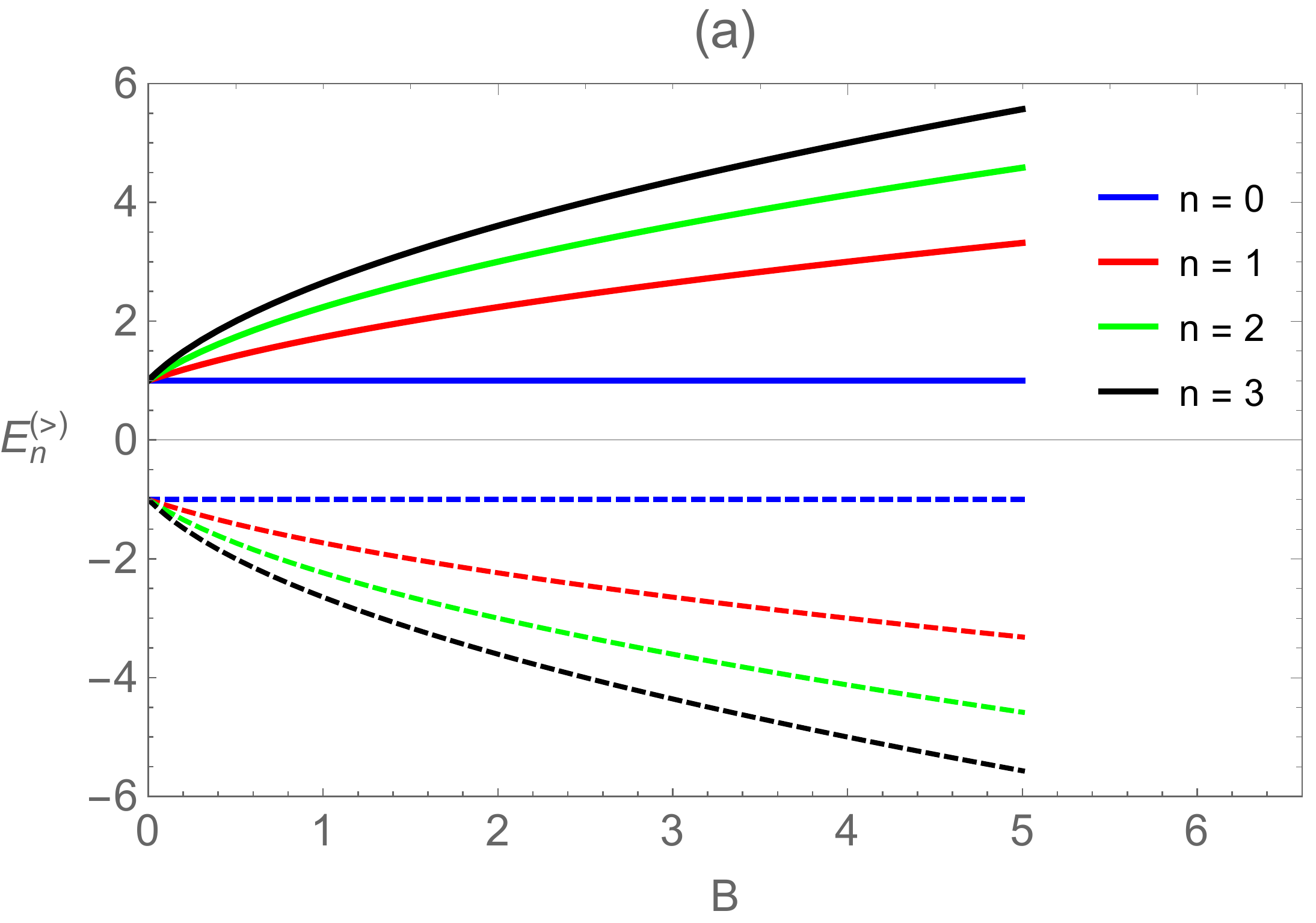}
	\includegraphics[width=\columnwidth]{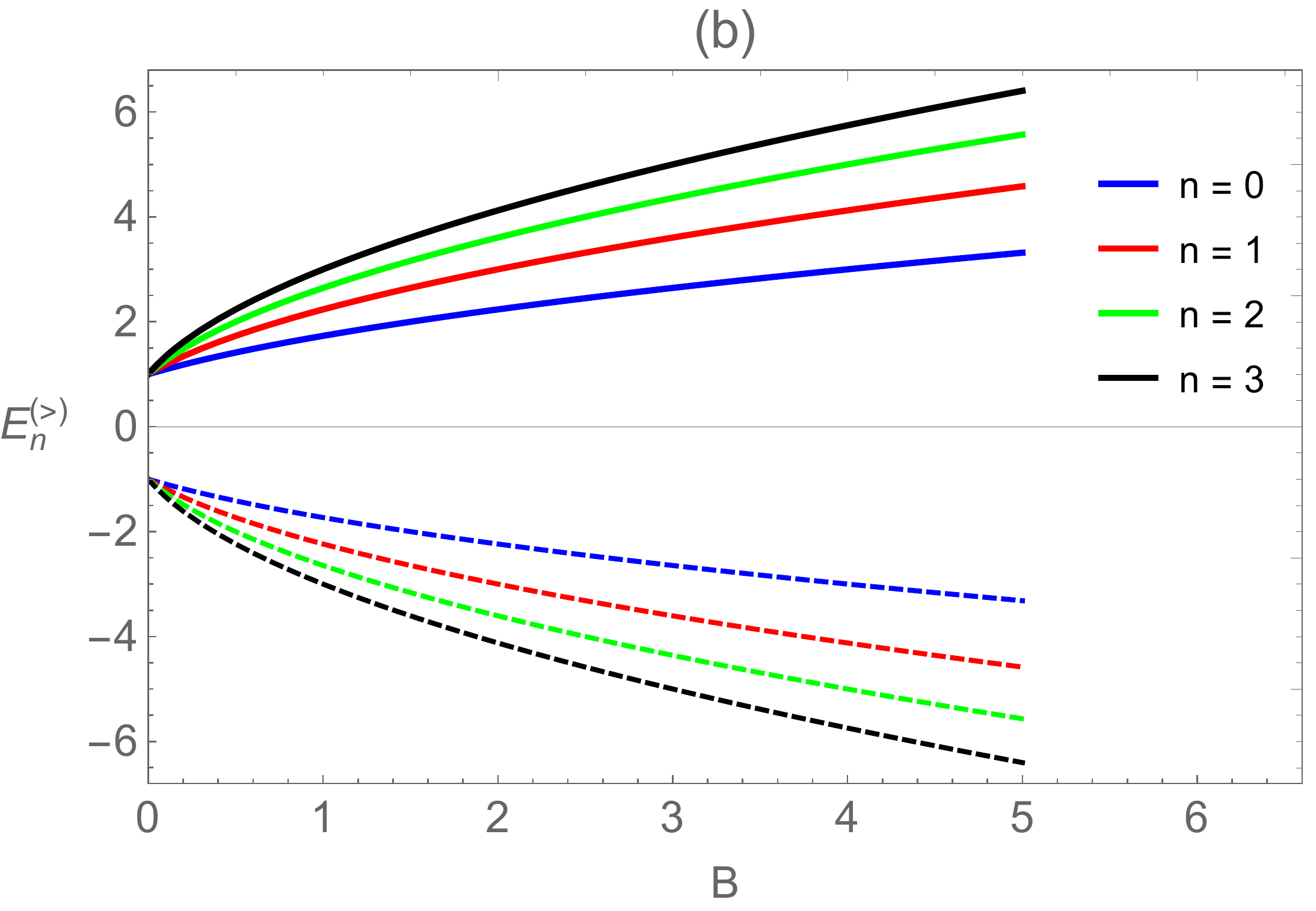}
    \caption{Sketch of the energy levels $E_{n}^{(>)}$ (Eq. (\ref{Ep})) as a function of the magnetic field $B$ for different values of $n$. In panel (a), $s=+1$ and in panel (b), $s=-1$. The positive energies are represented by solid lines and the negative by dashed lines. We assume $e=1$ and $M=1.$}
	\label{fig2DEb}
\end{figure}
Note that there are three other equivalent equations and there is no need to solve them here because their respective energies would also be equivalent. Equation (\ref{re}) is the confluent hypergeometric equation
 and its solution is well known. Thus, it can be shown that the solution to $\psi_a$ is \begin{align}
&\psi_{a}\left( r,\varphi \right) =c_{nm}\left( \frac{eB}{2}\right) ^{
\frac{1}{2}\left(1+{\frac{\left\vert L\right\vert }{\alpha}}\right)}e^{im\varphi }{r}^{{%
\frac{\left\vert L\right\vert }{\alpha }}}e{^{-\frac{1}{4}\,eB{r}^{2}}}
\notag \\
& \times {{_{1}F_{1}}\left(\frac{1}{2} \left(1+{\frac{\left\vert L\right\vert }{\alpha}}\right)+{\frac{{k}^{2}}{2eB}},\,1+{\frac{\left\vert L\right\vert }{\alpha
}},\,\frac{1}{2}eB{r}^{2}\right) }, \label{pphi}
\end{align}
where ${_{1}F_{1}}\left(a,b,z \right)$ denotes the confluent hypergeometric function of the first kind or Kummer's function $M(a,b,z)$ and $c_{nm}$ the normalization constant.
It can be shown that the hypergeometric function ${_{1}F_{1}}\left(a,b,z \right)$ has a divergent behavior for large values of $z$. Because of this, bound state solutions for Eq. (\ref{pphi}) are only possible if we impose that this function becomes a polynomial of degree $n$. For this to be accomplished, we require that $1/2+{\left\vert L\right\vert /2\alpha }+{{k}^{2}/2Be=-n}$, where $n\in\mathbb{Z}^{*}$, with $\mathbb{Z}^{*}$ denoting the set of the nonnegative integers.
\begin{figure}[!h]
	\centering	
	\includegraphics[width=\columnwidth]{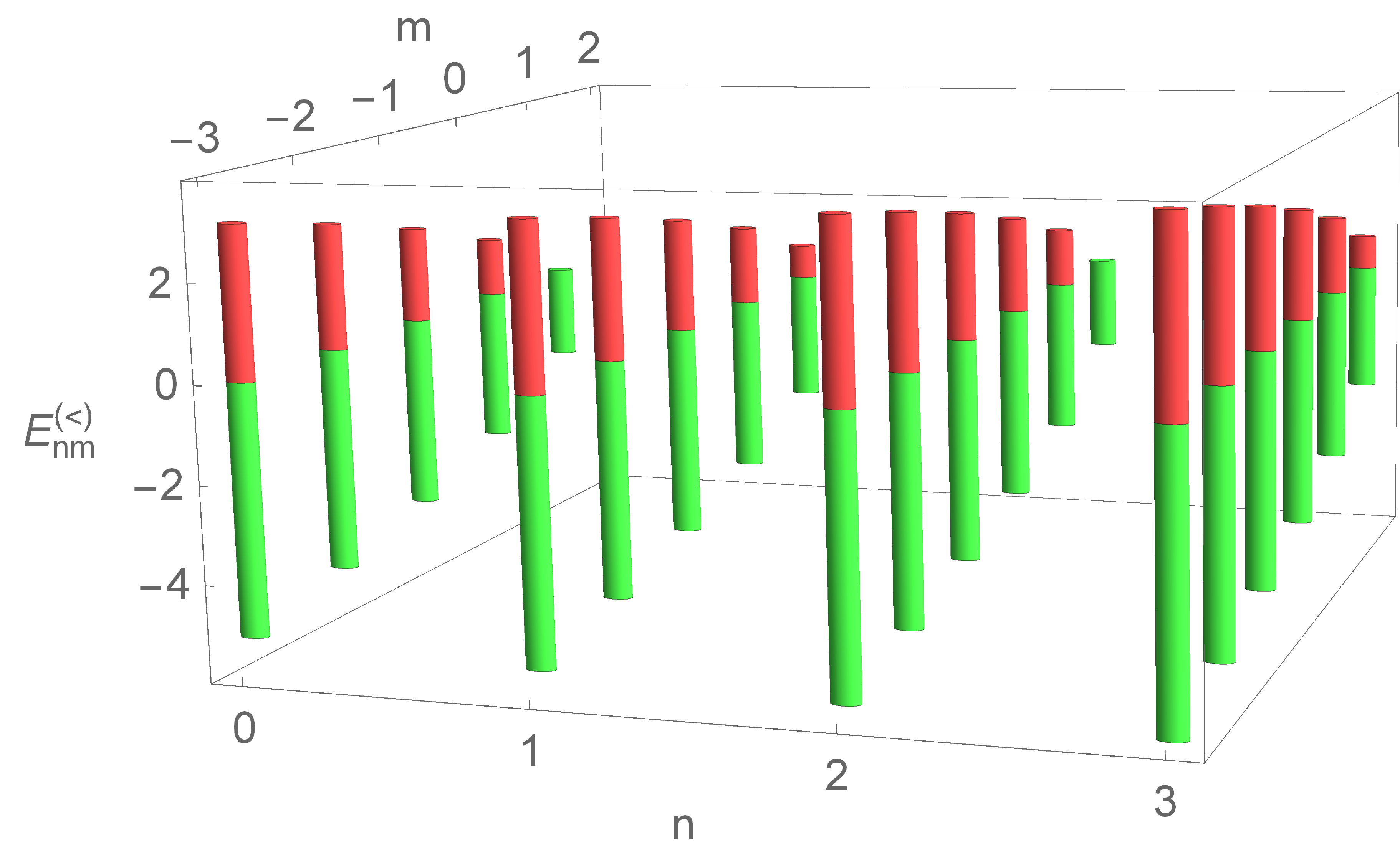}
    \caption{Sketch of the energy (Eq. (\ref{Em})) as a function of $n$ and $m$ for $a=0.5$, $\alpha =0.5$, $B=1$, $e=1$, $M=1$, $s=1$ and $\phi =1$.}
	\label{En3Dnm}
\end{figure}However, as we can see in Eq. (\ref{am}), the absolute value of the effective angular moment $L$ is defined in terms of the energy $E$. In this way, to obtain the energy eigenvalues from the above condition, we must consider $|L|>0$ and $|L|<0$, respectively, and then solve them for $E$. By making this, we get
\begin{align}
E_{n}^{(>)}& = \pm\sqrt{eB\left( 2n-s+1\right) +M^{2}}, \label{Ep}\\
E_{nm}^{(<)}& =-\frac{aeB}{\alpha } \pm \frac{1}{\alpha }\sqrt{
a^{2}e^{2}B^{2}+\alpha Q},\label{Em}
\end{align}
with the following requirement:
\begin{equation}
a^{2}e^{2}B^{2}+\alpha \,Q \geqslant 0, \label{cnd}
\end{equation}
where
\begin{equation*}
Q=\alpha eB\left( 2n-\frac{2}{\alpha }\left( m-\phi +\frac{s}{2}\right)
+1\right) +\alpha M^{2}.
\end{equation*}
\begin{figure}[!h]
	\centering	
	\includegraphics[width=\columnwidth]{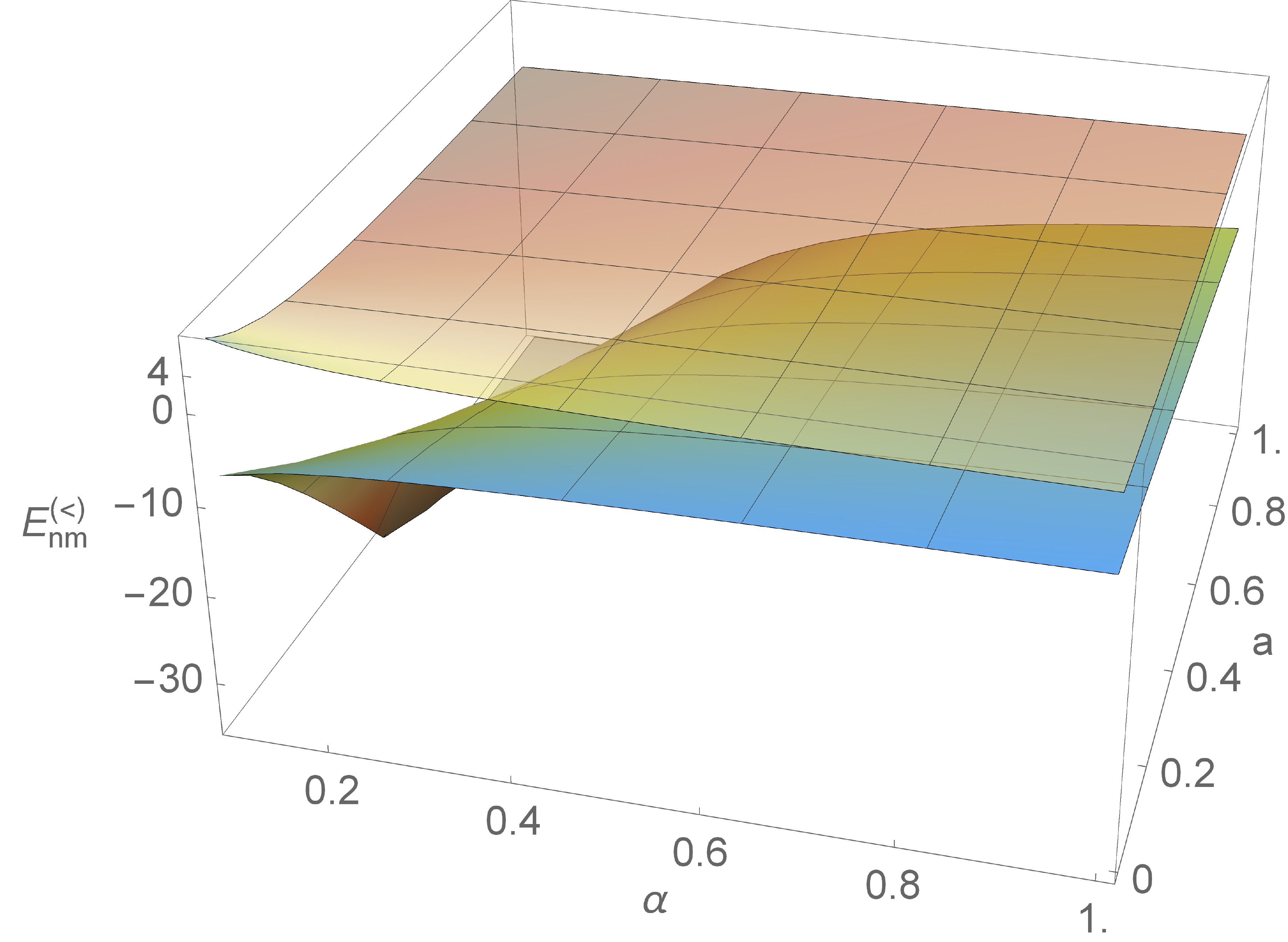}
    \caption{Sketch of the energy (Eq. (\ref{Em})) as a function of $\alpha$ and $a$ for $B = 4$, $M = 1$, $e=1$, $n=1$, $\phi=2$, $s = 1$ and $m = 1$.}
	\label{En3Daa}
\end{figure}
In Eqs. (\ref{Ep}) and (\ref{Em}), the superscripts ($>,<$) refer to the energies calculated for $\left\vert L\right\vert >0$ and $\left\vert L\right\vert <0$, respectively.
For a given choice of the element of spin $s$, the energy $E_{n}^{(>)}$ depends only on the quantum number $n$ and the magnetic field $B$. For a given value of $n$, the energy increases when the magnetic field is increased. In Fig. \ref{fig2DEb}, we show the profile of $E_{n}^{(>)}$ for the first four states for $s=1$. The energy levels for $s=-1$ (Fig. \ref{fig2DEb}(b)) are slightly larger than the profile for the case $s=1$ (Fig. \ref{fig2DEb}(a)).

The energies (\ref{Ep}) and (\ref{Em}) denote the relativistic Landau levels in the present context. These energies can be directly compared with those obtained for the relativistic oscillator (Dirac oscillator) addressed in Ref. \cite{EPJC.2019.79.311}. Although that scenario is different from the one we are exploring here, there are similarities between the profiles of the energy levels in both models. For example, for $s=1$, the energy (48) of the Ref. \cite{EPJC.2019.79.311} depends only on the frequency of the oscillator and the quantum number $n$. In our case, by defining the cyclotron frequency $\omega _{c}=eB/M$, Eq. (\ref{Ep}) results
\begin{equation}
\tilde{E}_{nm}^{(>)}=\pm \sqrt{2nM\omega _{c}+M^{2}},
\end{equation}
which makes such a similarity clear. Since the energies (\ref{Em}) are the only ones that depend on all the physical parameters involved in the current problem, we study them in more detail. For a given set of fixed parameters, for example, $a=0.5$, $\alpha =0.5$, $B=1.0$, $e=1.0$, $M=1$, $s=1$ and $\phi=1$, we have the profile of the energy levels as a function of $n$ and $m$ (Fig. \ref{En3Dnm}). We can clearly see that $|{E}_{nm}^{(<)}|$ increases with $n$ and $m$. The green solid bars denote the discrete energy values for a given $m$ and $n$.
\begin{figure}[!h]
	\centering	
	\includegraphics[width=\columnwidth]{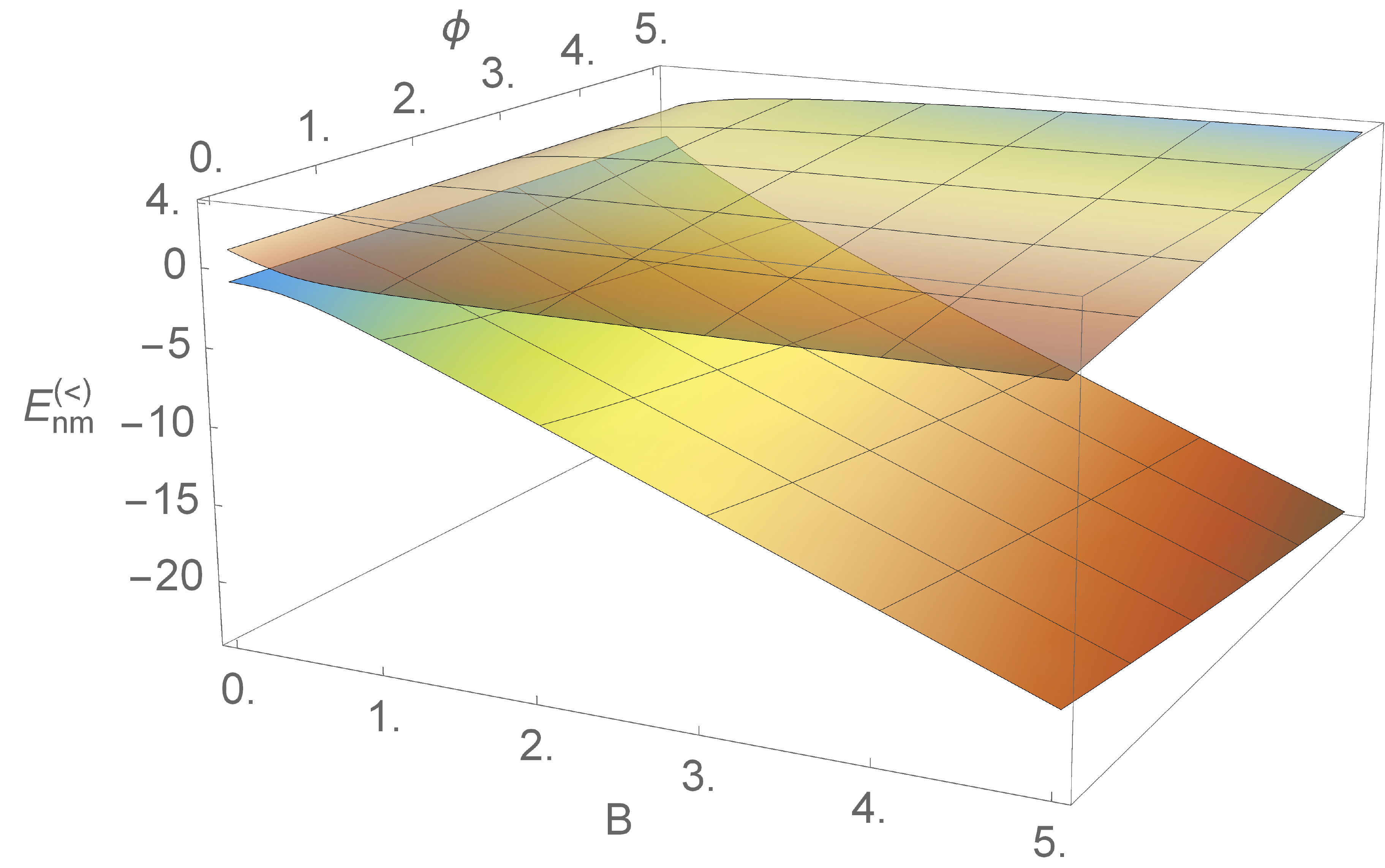}
    \caption{Sketch of the energy (Eq. (\ref{Em})) as a function of $B$ and $\phi$ for $a=1$, $\alpha =0.5$, $e=1$, $m=1$, $M=1$, $n=1$ and $s=1$.}
	\label{En3Dbf}
\end{figure}
On the other hand, when we investigate the behavior of (\ref{Em}) as a function of $\alpha$ and $a$ for specific values of the other parameters, we see that the negative spectrum changes more rapidly when compared with the positive one (Fig. \ref{En3Daa}). In the positive spectrum, both rotation and curvature lead to a linear change, except in the region with $\alpha < 0.3$ and arbitrary $a$ . In the negative spectrum, we see that the curvature effects are more predominant in the region where alpha has values smaller than 0.2. In this region, any variation in the rotation parameter implies in an abrupt change in the energy spectrum. Modifications in the energies with $\alpha < 0.3$ is an expected manifestation in our analyses. Its physical implication is inherent in the metric (\ref{metric}) and is an immediate consequence of the topological cone, which becomes more singular for smaller $\alpha$ values. To complete our analysis, we investigate the profile of the energy (\ref{Em}) as a function of magnetic field $B$ and the magnetic flux through the solenoid, $\phi$. Similarly to Fig. \ref{En3Daa}, by fixing the other parameters, we see that the energy of the anti-particle varies more rapidly when compared to the energy of the particle (Fig. \ref{En3Dbf}). Clearly, we observe that the energy of the particle varies very slowly throughout the region of flux and magnetic field . As a final commentary, we clarify that the cases discussed in Figs. \ref{En3Dnm}, \ref{En3Daa} and \ref{En3Dbf} can be investigated for other fixed parameter values. In this way, it can be shown that there are forbidden energies, depending on the values of the parameters considered. In general, this occurs when both the $\alpha$ parameter and the rotation parameter $a$ are smaller than $0.3$ and the other parameters assuming higher values than those we use here.

\section{Conclusions}
\label{sec:conclusions}

In the present manuscript, we have addressed the problem of the relativistic quantum motion of an electron in the spinning cosmic string background considering the presence of a uniform magnetic field and the Aharonov-Bohm potential. We have shown that this combination of potentials allows bound states configurations in the scenario of first-order solutions as well as in the case of second-order solutions of the Dirac equation.  It is worth noting the role played by the two different terms in the vector potential. As already known in the literature, we have shown that the uniform field is responsible for a behavior analog to a harmonic oscillator, which leads to the relativistic Landau quantization while the Aharonov-Bohm flux contributes to the angular momentum of the particle. In the case of first order solutions, which were obtained by solving Eqs. (\ref{fnl}) and (\ref{slem}) for $E=+M$ and $E=-M$, respectively, the oscillator-like behavior provided by the uniform magnetic field guarantees the convergent first-order solutions and, consequently, the existence of bound states. The isolated solutions obtained (Eqs. (\ref{fnl}) and (\ref{slem})) are particular solutions of the Dirac equation (\ref{diracsc}).

We have also studied the more general problem by solving the second-order equation implied by equations (\ref{Eqpsi}) and (\ref{Eqchi}) for the upper component of the Dirac spinor for $E\neq \pm M$. Using appropriate solutions (Eq. (\ref{S1})) we have derived the radial equation and shown that its solution is given in terms of the Kummer functions from which we have extracted the expression for the energy levels of the particle (Eqs. (\ref{Ep}) and (\ref{Em})). For the field configuration considered, we have found that the effective angular momentum of the electron depends on its energy and the Aharonov-Bohm flux tube while the potential vector that generates the uniform field leads to a charged oscillator. This implies that such field superposition provides distinct effects on the motion of the particle. Additionally, in some cases, the rotation produces a combined effect with both the uniform magnetic field and the curvature (see Eq. (\ref{Em})). We have shown that the energy levels of the particle and antiparticle depend on the values of the physical parameters involved. In the case of energy (\ref{Em}), its validity is conditioned to Eq. (\ref{cnd}). Depending on the choice we make for the parameters, we can obtain forbidden energies. The sketches in Figs. \ref{En3Dnm}, \ref{En3Daa}, and \ref{En3Dbf}  illustrate the profiles of the particle and antiparticle energies and show that they belong to the same spectrum. The effects of curvature and rotation are more evident when $\alpha < 0.3$, being the antiparticle energy the most affected. As a final comment, we would like to emphasize that the model studied in this article generalizes others found in the literature, such as those of Refs. \cite{oliveira2019topological,wang2018study} for the case including a superposition of external magnetic fields and the investigation of isolated solutions of the Dirac equation. Furthermore, we present a detailed discussion on the energy levels of the particle which, in general, is not found in the literature.

\section*{Acknowledgments}
We would like to thanks E.R.B. Mello (Universidade Federal da Para\'{i}ba, PB, Brazil) for his remarks and comments.
This work was partially supported by the Brazilian agencies CAPES, CNPq and
FAPEMA. EOS acknowledges CNPq Grants 427214/2016-5 and 307203/2019-0, and
FAPEMA Grants 01852/14 and 01202/16. This study was financed in part by the Coordena\c{c}\~{a}o de
Aperfei\c{c}oamento de Pessoal de N\'{\i}vel Superior - Brasil (CAPES) -
Finance Code 001. MMC acknowledges CAPES
Grant 88887.358036/2019-00.

\bibliographystyle{apsrev4-2}

\end{document}